\documentclass[final
  ]
  {aipproc}

\layoutstyle{6x9}

\def\beq{\begin{equation}}
\def\eeq{\end{equation}}
\def\bea{\begin{eqnarray}}
\def\eea{\end{eqnarray}}

\def\mpT{p_T \hspace{-1em}/\;\:}

\usepackage{amsmath}   
\usepackage{amsfonts}  
\usepackage{amssymb}

\begin{document}

\title{Correlations between sneutrino-antisneutrino oscillations and signatures at the LHC in anomaly-mediated supersymmetry breaking}


\author{Dilip Kumar Ghosh}{
  address={Department of Theoretical Physics and Centre for Theoretical
Sciences, Indian Association for the Cultivation of Science, 2A $\&$ 2B Raja S.C. Mullick Road, Kolkata 700 032, India}
}

\author{\underline{Tuomas Honkavaara}}{
  address={Department of Physics, and Helsinki Institute of Physics, P.O.Box 64, FIN-00014 University of Helsinki, Finland},
}

\author{Katri Huitu}{
  address={Department of Physics, and Helsinki Institute of Physics, P.O.Box 64, FIN-00014 University of Helsinki, Finland}
}

\author{Sourov Roy}{
  address={Department of Theoretical Physics and Centre for Theoretical
Sciences, Indian Association for the Cultivation of Science, 2A $\&$ 2B Raja S.C. Mullick Road, Kolkata 700 032, India}
}

\begin{abstract}
Sneutrino-antisneutrino oscillation can be observed at the LHC by studying a charge asymmetry of the leptons in the final states. In this talk, we demonstrate this in the context of an anomaly-mediated supersymmetry breaking model which can give rise to a large oscillation probability. The preferred region of the parameter space is characterized by the presence of a sneutrino next-to-lightest supersymmetric particle and a stau lightest supersymmetric particle. We show that the signals studied here have certain correlations with the pattern of the sneutrino oscillation.
\end{abstract}

\maketitle

%%%%%%%%%%%%%%%%%%%%%%%%%%%%%%%%%%%%%%%%%%%%
%% MAINMATTER
%%%%%%%%%%%%%%%%%%%%%%%%%%%%%%%%%%%%%%%%%%%%

Sneutrino-antisneutrino mixing occurs in any supersymmetric (SUSY) model where neutrinos have nonzero Majorana masses. Such $\Delta L = 2$ Majorana neutrino mass terms can induce a mass splitting ($\Delta m_{\tilde \nu}$) between the physical states. The effect of this mass splitting is to induce sneutrino-antisneutrino oscillations {\cite{hirschetal,grossman-haber1}}. This can lead to the sneutrino decaying into a final state with a ``wrong-sign charged lepton,'' and the lepton number can be tagged in sneutrino decays by the charge of the final state lepton. In this talk which is based on Ref. \cite{snu-lhc}, we assume that the sneutrino flavor oscillation is absent and lepton flavor is conserved in the decay of sneutrino/antisneutrino.

As discussed in \cite{tuomas}, the probability of finding a wrong-sign charged lepton in the decay of a sneutrino should be the time-integrated one and is given by
\bea
P(\tilde \nu \rightarrow \ell^+) = {\frac {x^2_{\tilde \nu}}
{2(1+x^2_{\tilde \nu})}}  {\cal B}_{\tilde \nu^{\ast}}({\tilde \nu}^{\ast}
 \rightarrow \ell^+ X)
\label{eqn-prob},
\eea
where the quantity $x_{\tilde \nu}$ is defined as $ x_{\tilde \nu} \equiv \Delta m_{\tilde \nu}/\Gamma_{\tilde \nu} $, and ${\cal B}_{{\tilde \nu}^{\ast}}$ is the branching ratio for $\tilde \nu^* \rightarrow \ell^+$. This signal can be observed from the single production of a sneutrino at the LHC, provided $ x_{\tilde \nu} \sim 1$ and ${\cal B}_{{\tilde \nu}^{\ast}}$ is significant.

Evidently, the probability of the sneutrino-antisneutrino oscillation depends crucially on $\Delta m_{\tilde \nu}$ and $\Gamma_{\tilde \nu}$. If $m_\nu \sim 0.1$ eV, the radiative corrections to the $m_\nu$ induced by $\Delta m_{\tilde \nu}$ face the bound \cite{grossman-haber1} $\Delta m_{\tilde \nu}/m_\nu \lesssim \mathcal{O} (4\pi/\alpha)$, 
implying $\Delta m_{\tilde \nu} \lesssim 0.1$ keV. Thus, in order to get $x_{\tilde \nu} \sim 1$, one also needs the sneutrino decay width $\Gamma_{\tilde \nu}$ to be $\sim \Delta m_{\tilde \nu} $. Because of the smallness of $\Gamma_{\tilde \nu}$, the sneutrino's lifetime would be large enough for sneutrino oscillation to take place before its decay.

However, for a spectrum where ${\tilde \chi^0_1}$ is the lightest supersymmetric particle (LSP), $\Gamma_{\tilde \nu} $ would not be $\lesssim {\cal O}(1)$ keV because of the presence of two-body decays
$\tilde \nu \rightarrow \nu \tilde\chi^0$ and $\tilde \nu \rightarrow \ell^- \tilde \chi^+$. If, instead, the mass spectrum is such that
\bea
m_{{\tilde \tau}_1} < m_{\tilde \nu} < m_{\tilde \chi^0_1}, m_{\tilde
\chi^\pm_1},
\label{spectrum}
\eea
where the lighter stau (${\tilde \tau}_1$) is the LSP, these two-body
decay modes are forbidden and the three-body decay modes such as $\tilde \nu \rightarrow \ell^- {\tilde \tau}_1^+ \nu_\tau$ and $\tilde \nu \rightarrow \nu {\tilde \tau}_1^\pm \tau^\mp$ are the available ones. However, having ${\tilde \tau}_1$ as a stable charged particle is strongly disfavored by astrophysical grounds. This can be avoided, for example, if a very small $R$-parity violating coupling $ (\lesssim 10^{-8})$ induces the decay ${\tilde \tau}_1 \rightarrow \ell \nu$, which occurs outside the detector after producing a heavily ionized charged track in the detector.

\begin{figure}[tb]
%\centering
\includegraphics[height=5.00cm]{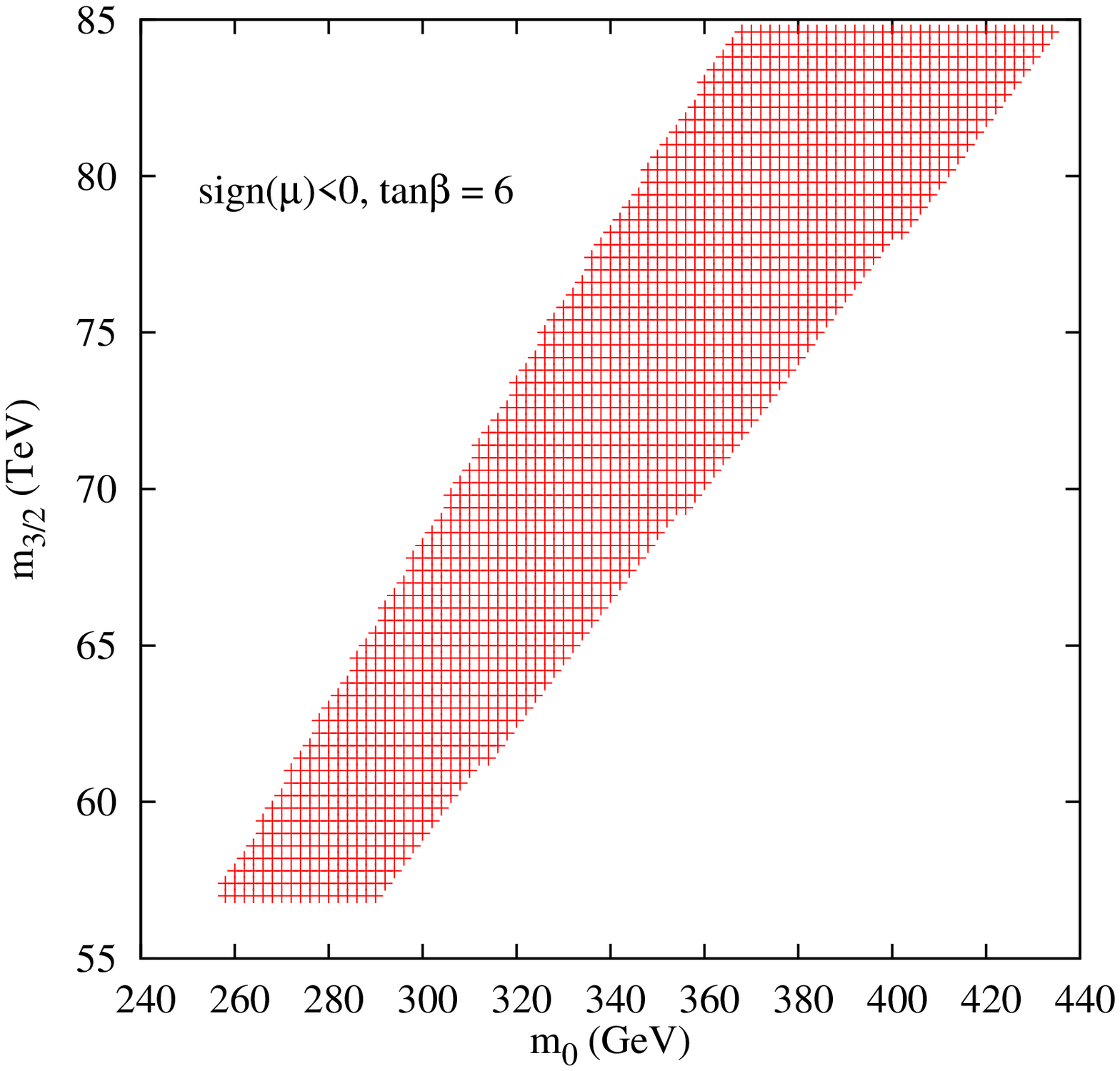}
\hspace{0.5cm}
\includegraphics[height=5.00cm]{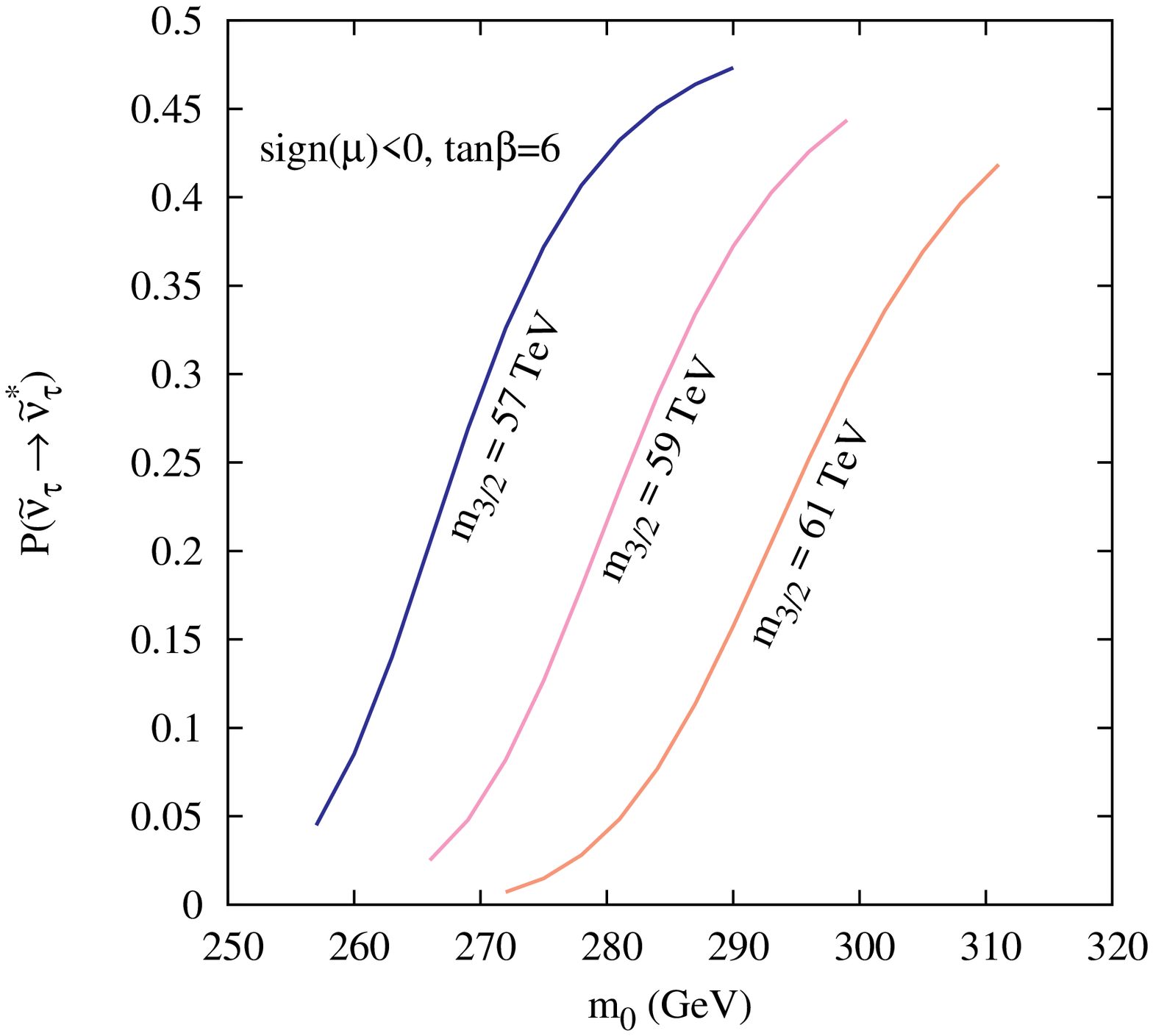}

\caption{Parameter space of the AMSB model with sneutrino NLSP and 
${\tilde \tau}_1$ LSP and $m_{\tilde{\ell}_{1,2}}<m_{\tilde{\chi}_1^0,\tilde{\chi}_1^\pm}$ ($\ell=e,\mu$) (on the left). ${\tilde \nu}_\tau$ oscillation probability as a function of $m_0$ (on the right).}
\label{amsb-para-space-osc-prob}
\end{figure}

The required spectrum (\ref{spectrum}) can be obtained in some region of the anomaly-mediated supersymmetry breaking (AMSB) \cite{amsb} parameter space with $\Delta m_{\tilde \nu} \lesssim \mathcal{O} (4\pi m_\nu/\alpha)$. In our analysis, we have $m_{\nu_i} \lesssim 0.3$ eV ($i=e,\mu,\tau$). In Fig. \ref{amsb-para-space-osc-prob} (on the left side), we display the region of the parameter space in $m_0 - m_{3/2}$ plane with ${\rm sign}(\mu) < 0 $ and $\tan\beta = 6$, where the above spectrum is valid. In this parameter space, $m_{\tilde{\ell}_{1,2}}<m_{\tilde{\chi}_1^0,\tilde{\chi}_1^\pm}$ ($\ell=e,\mu$).

In Fig. \ref{amsb-para-space-osc-prob} (on the right side), we plot the ${\tilde \nu}_\tau$ oscillation probability as a function of the common scalar mass $m_0$ for three different choices of $m_{3/2}$ with ${\rm sign}(\mu) < 0 $ and $\tan\beta = 6$ in the allowed parameter space. This figure tells that the probability of oscillation can be quite high. Hence, the AMSB has a good potential to produce signals of sneutrino-antisneutrino oscillation, which can be tested in colliders (noticed also earlier; see \cite{tuomas,like-sign-norp}).

%%%%%%%%%%%%%%%%%%%%%%%%%%%%%%%%%%%%%%%%%%%%%%%%%%%%%%%%%%%%%%%%%%%%%%%
%%%%%%%%%%%%%%%%%%%%%%%%%%%%%%%%%%%%%%%%%%%%%%%%%%%%%%%%%%%%%%%%%%%%%%%

In this work, the first production process we will consider is  
\begin{eqnarray}
p p  \to {\tilde \nu}_\tau {\tilde \tau}_1^+
\label{ppstau1}.
\end{eqnarray}
Since ${\tilde \nu}_\tau$ decaying to a three-body final state with $\tau^\pm$ is difficult to identify, we look at other channels mediated by virtual $W^-$ and $H^-$. If the ${\tilde \nu}_\tau$ oscillates into a ${\tilde \nu}_\tau^*$, we can have a three-body final state, ${\tilde \nu}_\tau \rightarrow {\tilde \nu}_\tau^* \rightarrow \ell^- {\tilde \tau}_1^+ {\bar \nu}_{\ell}$ leading to $\ell^- {\tilde \tau}_1^+ {\tilde \tau}_1^+ + \mpT \,$ signature from the process in Eq. (\ref{ppstau1}). Here, $\ell= e,~\mu$. The cross section for this process 
is given by $\sigma_\mathrm{osc}= \sigma(p p  \rightarrow {\tilde \nu}_\tau {\tilde \tau}_1^+) \times P_{{\tilde \nu}_\tau \rightarrow 
{\tilde \nu}_\tau^*} \times {\cal B}_{\tilde \nu^\ast}
({\tilde \nu}_\tau^* \rightarrow
\ell^- {\tilde \tau}_1^+ {\bar \nu}_{\ell}) $, where 
$P_{{\tilde \nu}_\tau \rightarrow {\tilde \nu}_\tau^\ast}$ denotes the 
sneutrino oscillation probability. When the ${\tilde \nu}_\tau$ survives as a ${\tilde \nu}_\tau$, one of the possible three-body decays of the ${\tilde \nu}_\tau$ is ${\tilde \nu}_\tau \rightarrow \ell^+ {\tilde \tau}_1^- \nu_{\ell}$. This would lead to $\ell^+ {\tilde \tau}_1^- {\tilde \tau}_1^+ + \mpT $ signature from the same process (\ref{ppstau1}). From these oscillation and no oscillation signals, one can define a charge asymmetry parameter
\begin{eqnarray}
A_\mathrm{asym} \equiv {{\sigma (\ell^-\tilde\tau_1^+\tilde\tau_1^+ +\mpT)
-\sigma (\ell^+\tilde\tau_1^-\tilde\tau_1^+ +\mpT)}\over
{\sigma (\ell^-\tilde\tau_1^+\tilde\tau_1^+ +\mpT)
+\sigma (\ell^+\tilde\tau_1^-\tilde\tau_1^+ +\mpT)}}.
\label{asymosc}
\end{eqnarray}
Since ${\cal B}_{\tilde \nu}({\tilde \nu}_\tau 
\rightarrow \ell^+ {\tilde \tau}_1^- 
{\nu_\ell}) = {\cal B}_{\tilde \nu^\ast} 
({\tilde \nu}_\tau^* \rightarrow \ell^- {\tilde \tau}_1^+ 
{\bar \nu}_{\ell}) $, one can rewrite Eq. (\ref{asymosc}) 
in the form $A_\mathrm{asym} = P_{{\tilde \nu}_\tau \rightarrow {\tilde \nu}_\tau^*} -P_{{\tilde \nu}_\tau \rightarrow {\tilde \nu}_\tau}$, where $P_{{\tilde \nu}_\tau \rightarrow {\tilde \nu}_\tau }$ denotes the sneutrino survival probability. From this, it is evident that $A_\mathrm{asym} = -1$ corresponds to no sneutrino oscillation. Hence, any deviation of $A_\mathrm{asym}$ from $-1$ is the smoking gun signature of sneutrino oscillation.

There is very little SM background to these signals assuming that the long-lived staus produce heavily ionized charged tracks which can be distinguished from the muon tracks. This is possible, since the staus are much slower than the muons because of their large masses. However, there are several other SUSY processes which can give rise 
to the same final state as our signal. These processes are
\bea
&&pp\rightarrow \tilde\nu_\ell \tilde \ell_L^+ \,\, {\rm with}\,\, \ell=e,\mu, \quad {\rm and}\quad
pp\rightarrow \tilde\chi_1^0\tilde\chi_1^+.
\eea
The relevant decay modes for these SUSY backgrounds with an example of different cross sections for a certain parameter point are presented in detail in \cite{snu-lhc}. All these backgrounds need to be considerd when calculating the asymmetry \eqref{asymosc}.

We select the signal events with the following criteria$\colon$ 
1) $p^{\ell^\pm}_T> 5$ GeV, 2) $|\eta^{{\ell^\pm},{\tilde \tau}_1}|< 2.5$, 3) transverse momentum of both ${\tilde \tau}_1^-$'s must satisfy $p^{{\tilde \tau}_1}_T> 100$ GeV, and 4) $\mpT < 20$ GeV. The last two cuts are crucial in clearly identifying signals from the SUSY background.

In Table \ref{table1}, we show the asymmetries including the SUSY background for three different parameter choices. In all of these cases, the oscillation probability is more than $0.15$. It is seen from the Table that, already with 30 fb$^{-1}$, one can distinguish between the oscillation and no oscillation cases in these sample points. When $\tan\beta$ grows, the ratio between the SUSY signal and the background reduces. Thus, this measurement, with the cuts used, is possible for small $\tan\beta$.

%%%%%%%
\begin{table}[tb]
%\begin{center}
\centering
\footnotesize
\begin{tabular}{|l|c|c|c|c|}
\hline
Parameter point & $A_\mathrm{asym}$ & 
\multicolumn{3}{c|}{$\pm$ Errors} \\
\cline{3-5}
$\tan\beta$, $m_0$(GeV), &osc. &  &  & \\
$m_{3/2}$(TeV) &(no osc.)& 30 fb$^{-1}$ & 100 fb$^{-1}$& 300 fb$^{-1}$ \\
\hline
5, 370, 81, $\mu < 0 $ & -0.515 & 0.072 & 0.040 &0.023 \\
& (-0.859) & (0.043) & (0.024) & (0.014)
\\
\hline
6, 270, 57, $\mu < 0 $ & -0.325 & 0.052 & 0.029 & 0.017 \\
& (-0.676) & (0.041) & (0.022) & (0.013)
\\
\hline
7, 248, 49, $\mu < 0 $ & -0.149 & 0.044 & 0.024 & 0.014 \\
& (-0.266) & (0.043) & (0.024) & (0.014)
\\
\hline
\end{tabular}
%\end{center}
\caption{Asymmetries and the corresponding errors for different parameter points. Numbers in the brackets are for the no oscillation case.} 
\label{table1}
\end{table}
%%%%%%%

\begin{figure}[tb]
\centering
\includegraphics[height=5.00cm]{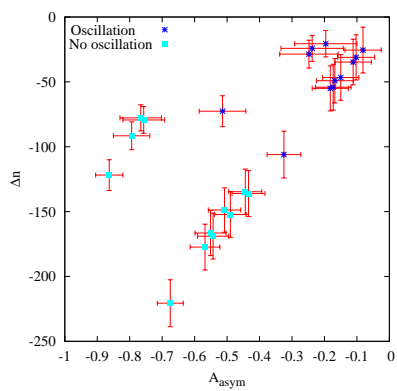}
\caption{Correlation between $\Delta n$ and $A_\mathrm{asym}$ for the
oscillation and the no-oscillation cases for different parameter points.}
\label{correlation}
\end{figure}

If the SUSY spectrum is not known, one can still deduce in favorable
cases whether there is sneutrino oscillation or not. We demonstrate this in Fig.~\ref{correlation} for $\tan\beta=5,6$ and the values of $m_0$ and $m_{3/2}$ for which the signal cross sections are large. Here, it has been required that oscillation probability is more than
0.25 and $S/\sqrt{B} \gtrsim 5 $. We plot the difference ($\Delta n$) between the numbers of events for $pp \to \ell^- {\tilde \tau}_1^+ {\tilde \tau}_1^+ + \mpT$ and $pp \to \ell^+ {\tilde \tau}_1^- {\tilde \tau}_1^+ + \mpT$ for integrated luminosity 30 fb$^{-1}$ vs the asymmetry. The corresponding errors are shown at the $1 \sigma$ level.
%%%%%%%%%%%%%%%%%%%%%%%%%%%%%%%%%%%%%%%%%%%%%%%%%%%%%%%%%%%%%%%%%%%%%%%
%%%%%%%%%%%%%%%%%%%%%%%%%%%%%%%%%%%%%%%%%%%%%%%%%%%%%%%%%%%%%%%%%%%%%%%
One can see from this correlation plot that the sneutrino oscillation 
represents bigger asymmetry and bigger $\Delta n$, whereas, in the
case of no sneutrino oscillation, the value of $\Delta n$ and the 
asymmetry should be on the smaller side. This is expected, since, with the cuts that we have imposed, the ${\tilde \nu}_\tau$ type of oscillation signal is stronger. When there is oscillation, the splitting between two different charge final states is smaller, and, naturally, the asymmetry is closer to zero.

In this study, we have assumed that the staus decay outside the detector. It is also possible that the $R$-parity violating coupling is larger and the staus decay inside the detector after traversing a certain length or they decay promptly. We hope to come back to these issues in a future work \cite{sneutrino_next_work}. 

This work is supported in part by the Academy of Finland (Project No. 115032). TH thanks the V\"ais\"al\"a foundation for support.


\begin{thebibliography}{99}

\bibitem{hirschetal}
M. Hirsch, H.V. Klapdor-Kleingrothaus, and S.G. Kovalenko, Phys. Lett. B \textbf{398}, 311 (1997). arXiv:hep-ph/9701253.

\bibitem{grossman-haber1}
Y. Grossman and H.E. Haber, Phys. Rev. Lett. \textbf{78}, 3438 (1997). arXiv:hep-ph/9702421.

\bibitem{snu-lhc} D. Ghosh, T. Honkavaara, K. Huitu, and S. Roy, Phys. Rev., D \textbf{79}, 055005 (2009). arXiv:0810.0913 [hep-ph]. 

\bibitem{tuomas} T. Honkavaara, K. Huitu, and S. Roy, Phys. Rev. D \textbf{73}, 055011 (2006). arXiv:hep-ph/0512277.

\bibitem{amsb} G.F. Giudice, M.A. Luty, H. Murayama and R. Rattazzi, JHEP \textbf{9812}, 027 (1998). arXiv:hep-ph/9810442; L. Randall and R. Sundrum, Nucl.\ Phys.\  B \textbf{557}, 79 (1999). arXiv:hep-th/9810155.

\bibitem{like-sign-norp} K. Choi, K. Hwang, and W.Y. Song, Phys. Rev. Lett. \textbf{88}, 141801 (2002). arXiv:hep-ph/0108028.

\bibitem{sneutrino_next_work}
D.K. Ghosh, T. Honkavaara, K. Huitu, and S. Roy (work in progress).

\end{thebibliography}
\end{document}